\documentclass[aps,twocolumn,superscriptaddress,floatfix,nofootinbib]{revtex4-1}
    
\usepackage{graphicx,amsmath,amssymb,verbatim,color}
\usepackage{booktabs}
\usepackage{enumitem} 
\usepackage{comment}
\usepackage{soul}
\usepackage{setspace}
\usepackage[dvipsnames]{xcolor}
\usepackage{bm}
\usepackage[utf8]{inputenc}
\usepackage[colorlinks=true,citecolor=blue,linkcolor=blue,urlcolor=blue, backref=false,pdfborder={0 0 0}]{hyperref}
\usepackage{float}
\usepackage{multirow}
\usepackage{dcolumn}
\usepackage{textgreek}
\usepackage[title, titletoc]{appendix}

\newcommand{\Planck}{\textit{Planck}}
\newcommand{\LCDM}{$\Lambda$CDM}

\newcommand{\Hunit}{{\rm km}\ {\rm s}^{-1}\rm{Mpc}^{-1}}

\def\Dperp{${\cal D}^{\rm perp}$}
\def\Dpar{${\cal D}^{\rm par}$}
\def\ltsima{$\; \buildrel < \over \sim \;$}
\def\gtsima{$\; \buildrel > \over \sim \;$}
\def\simlt{\lower.5ex\hbox{\ltsima}}
\def\simgt{\lower.5ex\hbox{\gtsima}}
\makeatletter
\def\bstctlcite{\@ifnextchar[{\@bstctlcite}{\@bstctlcite[@auxout]}}
\def\@bstctlcite[#1]#2{\@bsphack \@for\@citeb:=#2\do{%
  \edef\@citeb{\expandafter\@firstofone\@citeb}%
  \if@filesw\immediate\write\csname #1\endcsname{\string\citation{\@citeb}}\fi}%
  \@esphack}
\makeatother

\begin{document}

\pagenumbering{arabic}

\title{A precise test of the $\Lambda$CDM cosmology with baryon acoustic oscillations and supernovae}

\author{George Efstathiou}
\email{gpe@ast.cam.ac.uk}
\affiliation{Kavli Institute for Cosmology \& Institute of Astronomy, Madingley Road, Cambridge, CB2 0HA, UK.}

\date{\today} 
\begin{abstract}
We test the \LCDM\ cosmology inferred from \Planck\ and the Atacama Cosmology Telescope (ACT) using updated baryon acoustic oscillation (BAO)  measurements from the Dark Energy Spectroscopic Instrument (DESI) combined with the new Unite and Dark Energy Survey (DES) Dovekie supernova compilations. We rotate the BAO distances into components perpendicular and parallel to the cosmic microwave background (CMB) degeneracy lines, which isolate a geometric test of \LCDM\ from  a measurement of the matter density $\omega_m$ at each BAO redshift. The perpendicular distances agree with the \Planck-ACT+SN  \LCDM\ predictions over the redshift range $0.51-2.33$ to extremely high precision.
For the lowest DESI redshift, $z = 0.295$, the consistency of the perpendicular distance with the \Planck-ACT+SN prediction is dependent on the SN sample.  The Unite sample differs from  the CMB prediction by  $\sim 2.2\sigma$, while the results for DESI-Dovekie depend on the fidelity of a small number of SN at redshifts $z < 0.1$ which were not observed by DES.
 Small differences between the  SN samples can be absorbed by the extra degrees of freedom  associated with  parameterizations of the equation of state $p/\rho c^2 = w(z)$ leading to  apparent evidence for evolving dark energy.
From both SN samples, the  parallel distances  prefer a lower $\omega_m$ than the \Planck-ACT \LCDM\ value by about $2 \sigma$. Although this is not a serious discrepancy, we argue that the CMB results may also be affected by small systematics, including the reionization optical depth, possible deviations from a power-law curvature fluctuation spectrum on large scales, and differences between CMB likelihoods. Each of these potential systematics can mimic evolving dark energy.   Until such systematics are better understood,  we contend that there is  no compelling case at this time either to reject \LCDM\ or to conclude that dark energy evolves.
 
\end{abstract}
\maketitle

The existence of dark energy in our Universe is one of the greatest unsolved problems in physics. The standard six parameter \LCDM\ cosmology,  favored by
measurements of the cosmic microwave background (CMB)  anisotropies,  models
dark energy as a cosmological constant with a time-independent equation of
state $w = -1$. However,  since so little
is known about dark energy there is a compelling case to test this assumption. 
Recent measurements of baryon acoustic oscillations (BAO) by the DESI collaboration \cite{DESI:2024, DESI:2025} have been interpreted as  evidence that dark energy  evolves, spawning a large number of papers exploring theoretical explanations. Clearly, such a violation of the \LCDM\ paradigm would have major implications for fundamental physics. It is therefore important that the DESI analysis is subjected to independent scrutiny.
 
The preference for evolving dark energy over \LCDM\ presented in \cite{DESI:2024, DESI:2025}  was found to be strongly dependent on the specific compilation of Type Ia supernovae (here abbreviated to SN)  used in combination with the BAO and CMB measurements. BAO and CMB measurements taken together give only a mild preference for evolving dark energy, in the region of $\sim 2-3\sigma$, depending on the exact choice of data sets and statistical analysis. This paper investigates the sensitivity of claims for evolving dark energy using new
SN compilations  \cite{Popovic:2026, Hoyt:2026, Camilleri:2026}. 

To achieve this goal, we extend the analysis presented in \cite{Efstathiou:2025} specifically designed to test the \LCDM\ cosmology inferred from  CMB anisotropies \cite{Planck:2018, Rosenberg:2022, SPT:2023, ACT:2025, SPT:2026}. We use 
DESI DR2 BAO measurements over the full redshift range ($z = 0.295-2.33$), updating the Ly$\alpha$ analysis to the full shape results reported in \cite{DESI:2026}.
We refer to this data set as DR2$^\alpha$. We combine these data with two new supernova compilations: the DES-Dovekie compilation \cite{Popovic:2026b, Popovic:2026} which is a revision of the Dark Energy Survey 5 year (DES5Y) SN sample \citep{Sanchez:2024, Vincenzi:2024, DES5Y:2024}  using improved photometric calibrations  and the larger Supernova-Unite compilation \cite{Camilleri:2026}  which combines the Pantheon+ catalogue \cite{Scolnic:2022, Brout:2022a, Brout:2022b} with DES5Y using  methodological improvements developed for Dovekie and refined host galaxy mass measurements \cite{Lee:2026}.  In addition to differing in size, the Unite catalogue contains more SN
at  redshifts $<0.1$ ($545$) from more sources and has greater redshift  overlap with DES5Y SN. In contrast, the Dovekie sample includes a small number of SN ($187$) at $z<0.1$ mostly from the Foundation survey \citep{Foley:2018} and is therefore sensitive to the relative calibration of these SN with the more distant DES5Y SN. Consistency  between Unite and Dovekie gives an indication of the level of systematics in SN data.

\begin{figure*}
    \centering
{\hspace{-0.4truein}\includegraphics[width=0.98\textwidth]{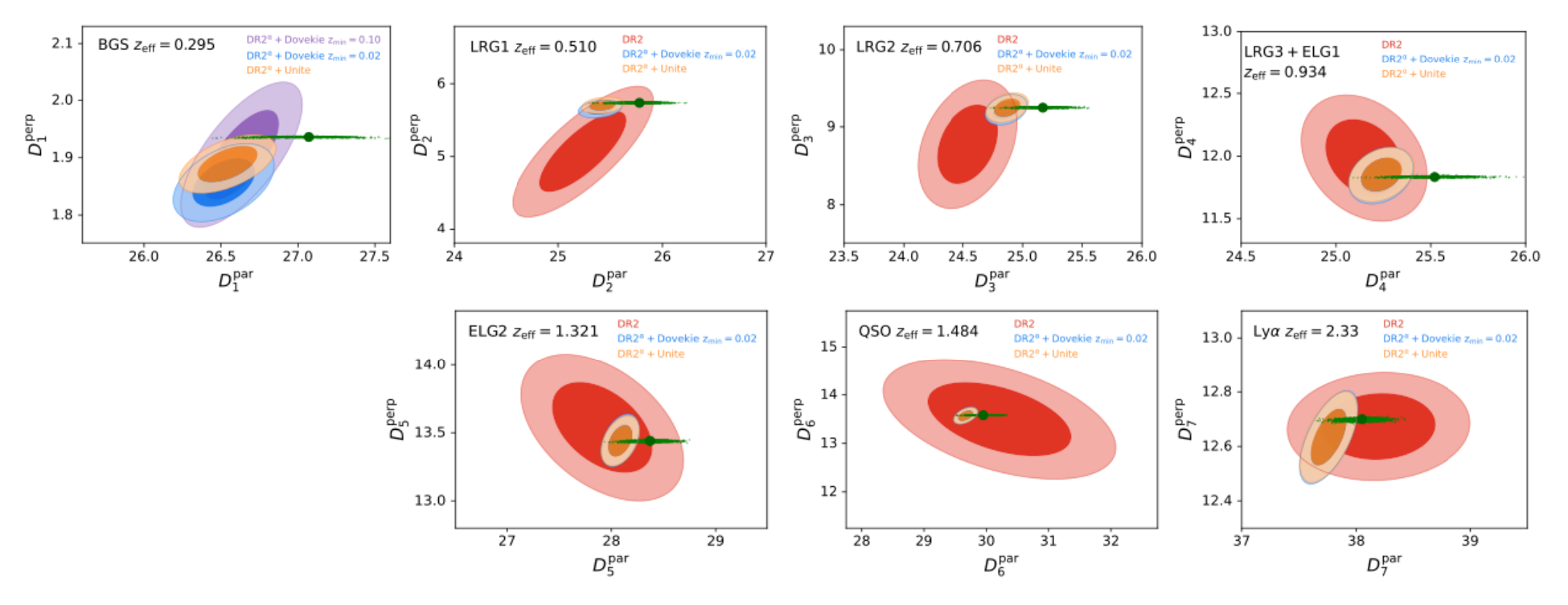}}
    \caption{The red $1\sigma$ and $2\sigma$ contours show the parameters \Dperp\ and \Dpar\ inferred from the DESI DR2 \cite{DESI:2025} galaxy, quasar and Ly$\alpha$ samples (using their notation). The dark green filled circles show the best fit from the P-ACT CMB likelihood used in this paper and the green points show samples from the CMB chains illustrating the CMB degeneracy direction and spread arising from the variation of $\omega_m$. The orange and blue contours show the  combination of DESI $^\alpha$+Unite and  DESI DR2$^\alpha$+ Dovekie respectively using SN in the redshift range $0.02< z < 1.0$ with $H(z)$ parameterized by  Eq. \ref{equ:Hz3}. The purple contour in the BGS panel shows  constraints eliminating  Dovekie SN  with $z< 0.1$. At higher redshifts,  the purple, orange and blue contours are almost indistinguishable.}
    \label{fig:Main_fig}
\end{figure*}

BAO surveys measure the angular diameter and Hubble distance ratios $D_{\rm M}(z)/r_d$ and $D_H(z)/r_d$, where
\begin{equation}
D_{\rm M} (z) =  \int_0^z dz^\prime {c\over H(z^\prime)},  \qquad
D_H(z)  =   {c \over H(z)}, \label{equ:BAO1}
\end{equation}
 $H(z)$ is the Hubble parameter and $r_d$ is the sound horizon. In \LCDM\ the Hubble parameter at low redshifts can be written as 
\begin{equation}
h(z) = \left [ \omega_m (1+z)^3 + (\alpha^2-\omega_m^3)/\omega_m^2\right]^{1/2}, \label{equ:Hz2}
\end{equation}
where $h(z)$ is the Hubble parameter in units of $100\ \Hunit$, $\omega_m = \Omega_mh^2$, $\alpha = \Omega_mh^3$ and $h \equiv h(z=0)$\footnote{We assume a spatially flat Universe throughout this paper.}. The parameter $\alpha$ is a proxy for
the acoustic peak location parameter $\theta_*$ and is determined to high accuracy
by the CMB ($\alpha = 0.09569 \pm 0.00043$ for the CMB data combination used in this paper). Eq. \ref{equ:Hz2} defines a line-like degeneracy in the space of $D_M-D_H$ with an extent  given by the error in $\omega_m$ determined from the CMB. For the accuracy
required in this paper we fit the slopes and $\omega_m$ dependence of the degeneracy directions directly to $D_M(z)/r_d$ and $D_H(z)/r_d$ at the relevant redshifts from the CMB MCMC chains, which accounts for the  weak dependence of $r_d$ on cosmological parameters.

The natural basis for comparing  BAO measurements to the predictions of \LCDM\
is to rotate the distances by an angle $\gamma(z)$,
\begin{subequations}
\begin{eqnarray}
{\cal D}^{\rm par} &=& (D_{H}/r_d) \sin \gamma(z) + (D_M/r_d) \cos \gamma(z), \label{equ:rot1a} \\
{\cal D}^{\rm perp} &=& (D_{H}/r_d) \cos \gamma(z) - (D_M/r_d) \sin \gamma(z), \label{equ:rot1b}
\end{eqnarray}
\end{subequations}
so that the CMB degeneracy line is parallel to the \Dpar\  axis \cite{Efstathiou:2025}. This rotation is not trivial: the parameter \Dperp\ provides a geometrical test of the \LCDM\ model in which the CMB prediction
has negligible error, while the parameter \Dpar\  provides a test of the CMB prediction for the parameter $\omega_m$.\footnote{The fact that $\omega_m$ is a dominant parameter in comparing BAO and \LCDM\ has been discussed  by Zaldarriaga \cite{Zaldarriaga:2026}.}

To compare with the CMB,  we use the P-ACT likelihood chains of \citep{ACT:2025}\footnote{Downloaded from https://lambda.gsfc.nasa.gov/product/act/\newline act\_dr6.02/act\_dr6.02\_chains\_lcdm\_get.html/~p{$-$}actbase\_lcdm\_camb.}
combining \Planck\ and ACT DR6 temperature and polarization power spectra, using the \Planck\ Commander TT-likelihood \cite{Plancklikelihood_2018} and SROLL2 EE-likelihood \citep{Delouis:2019,Pagano:2020} over the  multipole range $\ell=2-29$. The P-ACT likelihood constrains the reionization optical depth to $\tau = 0.060^{+0.0055}_{-0.0065}$ for the \LCDM\ cosmology. Note that the P-ACT likelihood gives closely similar results to the \Planck\ likelihood
from \cite{Rosenberg:2022} used in \cite{Efstathiou:2025}.

The red contours in Fig.~\ref{fig:Main_fig} show the constraints on \Dpar\ and \Dperp\ for the DESI DR2 sample.  Note that DESI DR2 were not able to measure $D_M/r_d$ and $D_H/r_d$ separately for
the nearby Bright Galaxy Sample (BGS), but instead quote a measurement of the isotropized 
distance $D_V/r_d = (zD^2_MD_H)^{1/3}/r_d$. For reference DESI DR2 measure
$D_V(z=0.295)/r_d = 7.942 \pm 0.075$ which is within $\sim 1 \sigma$ of the P-ACT
\LCDM\footnote{The necessity to refer specifically to the P-ACT \LCDM\ model will become clear towards the end of this paper.} expectation of $D_V(z=0.295)/r_d = 8.037\pm 0.057$. Figure~\ref{fig:Main_fig} is consistent with the results of \cite{Efstathiou:2025} who concluded that the \LCDM\ model was in good agreement with the DESI DR2 BAO measurements apart from a modest tension with \Dpar\ for the LRG2 sample. 

Type Ia SN provide a geometrical measurement of the luminosity distance $D_L(z) = (1+z)D_M(z)$. The combination of BAO and SN  can therefore be used to reconstruct $H(z)$. Such a reconstruction  is essentially theory free, depending only on the  metric of General Relativity, provided that the fitting function for $H(z)$ is a satisfactory fit to the data.   

The DESI collaboration fit their data to an evolving equation of state modeled by the CPL \cite{Chevallier:2001, Linder:2003} parameterization $w(z) = w_0 + w_a(z/(1+z))$. In this model, $h(z)$ varies as 
\begin{eqnarray}
h(z) & = &  \Bigg[\tilde\Omega_m(1+z)^{3}+(1-\tilde\Omega_m)(1+z)^{3(1+\tilde w_{0}+\tilde w_{a})}  \nonumber \\ 
 & & \times \left.\exp \left(-3\tilde w_{a}\frac{z}{1+z}\right)\right]^{1/2}.  \label{equ:Hz3} 
\end{eqnarray}
We can view Eq.~\ref{equ:Hz3} as a flexible smooth fitting function, with $\tilde\Omega_m$, $\tilde w_0$ and $\tilde w_a$ as additional parameters that may, or may not,   be related to the physical parameters $\Omega_m, w_0, w_a$.  We can then use the DESI DR2$^\alpha$ data, including the BGS measurement of $D_V$, in combination with SN data to fit\footnote{We use the
{\tt MULTINEST} nested sampler \citep{Feroz:2009, Feroz:2011}.} Eq.~\ref{equ:Hz3} and to predict
$D_M(z)$ and $D_H(z)$ at the redshifts of the DESI measurements. 

\begin{figure}
    \centering
{\hspace{0.1truein}\includegraphics[width=0.5\textwidth]{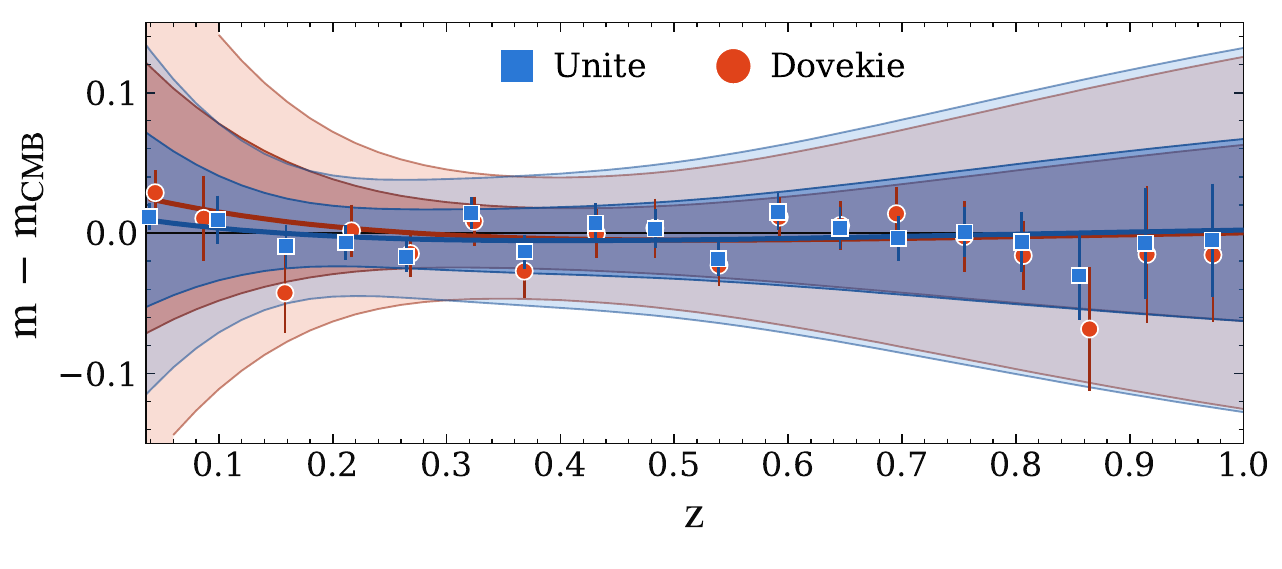}}
    \caption{Magnitude residuals of  Unite (blue) and Dovekie (red) SN relative to the best fit  P-ACT \LCDM\ cosmology. The colored lines show the best fit of Eq.~\ref{equ:Hz3} to SN  combined with the DESI DR2$^\alpha$ BAO measurements. Points  with $1\sigma$ errors show
    maximum likelihood band averages of the SN magnitudes computed as described
    in \cite{Efstathiou:2025b}.  The shaded regions show the $1\sigma$ and $2\sigma$ error ranges allowed by the fits.}
    \label{fig:Magres}
\end{figure}

Figure \ref{fig:Magres} shows the best fits to the magnitude residuals relative
to the P-ACT \LCDM\ model for both SN samples. Here
we use SN over the Hubble distance redshift range $0.02< z^{\rm HD} <1$, (Henceforth, we drop the superscript on $z^{\rm HD}$.) The P-ACT \LCDM\ cosmology provides an excellent  fit over the entire redshift range for the Unite sample, while the Dovekie sample shows a slight upturn at redshifts  $z \simlt 0.1$. 

\begin{figure*}
    \centering
{\hspace{-0.4truein}\includegraphics[width=0.98\textwidth]{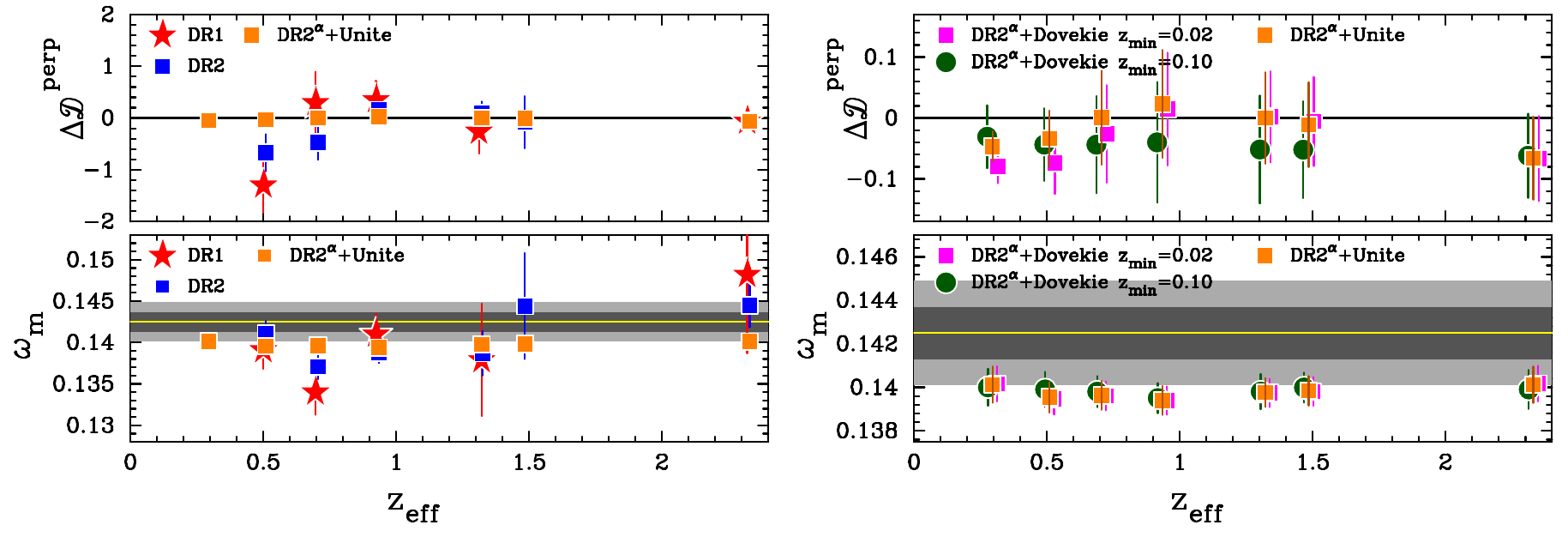}}
    \caption{The upper figures show residuals of the perpendicular distances, \Dperp,  relative to the P-ACT \LCDM\ expectations (see Table~\ref{tab:perp} in the Appendix). The lower panels show estimates of $\omega_m$ inferred from the parallel distances,  \Dpar,  (see Table~\ref{tab:parallel}). The grey bands show the $1\sigma$ and $2\sigma$ limits on $\omega_m$ inferred from the P-ACT  \LCDM\ cosmology. Figures to the left show results for DESI DR1, DESI DR2 and DR2$^\alpha$+Unite. Figures to the right show  DR2$^\alpha$+SN
    results on a greatly expanded scale.}
    \label{fig:Paper_fig2}
\end{figure*}

The orange and blue contours in Fig.~\ref{fig:Main_fig} show the constraints on \Dperp\ and
\Dpar\ inferred from these fits. The Appendix lists values for these parameters and for the additional parameters used to construct
Fig.~\ref{fig:Paper_fig2}. We first comment on the qualitative features of
these plots before discussing more detailed statistics. The results of Fig.~\ref{fig:Main_fig} are striking in several respects. Firstly, adding SN to DESI
BAO leads to a large reduction in the distance errors which is why inferences on evolving dark energy are so sensitive to small systematic errors in the SN data. Secondly, in every case except for the lowest redshift bin, the \Dperp\ contours intersect
the P-ACT \LCDM\ degeneracy lines. Thus, the geometrical constraints are consistent, to high precision, with the expectations of the P-ACT \LCDM\ cosmology over the redshift range
$0.510 - \ 2.33$. Thirdly the best fit P-ACT \LCDM\ points (shown by the green dots) always lie to the right of the BAO+SN contours. Visually, one can see that adjusting  just one parameter in the \LCDM\ cosmology, $\omega_m$,  the CMB and BAO+SN can be brought into near perfect agreement. Furthermore,  the
samples from the P-ACT \LCDM\ chains overlap with  the BAO+SN contours, so such a shift in  $\omega_m$ does not seem statistically implausible (this  will be quantified below). There is therefore no need for a more complex 
model to establish consistency between CMB, BAO and SN data over the redshift range 
$0.510 \ \text{---} \ 2.33$.

The  BAO+SN constraints at  the redshift of the BGS survey
in Fig.~\ref{fig:Main_fig} depend on the SN sample. \Dperp\ differs from the  P-ACT \LCDM\  value by respectively  $2.2\sigma$, $2.9\sigma$ and $0.6\sigma$ for  BAO+Unite, BAO+Dovekie ($z_{\rm min} = 0.02$) and BAO+Dovekie ($z_{\rm min} =0.1$).
 The trends evident  in both Fig.~\ref{fig:Magres} and Fig.~\ref{fig:Main_fig} are consistent with a small residual  systematic error in matching the photometry of DES5Y SN with low redshift SN in the Dovekie
 sample, though with a lower amplitude than previously suggested for the original DES5Y sample \cite{Efstathiou:2025b, Efstathiou:2025, Notari:2025} (see also \cite{Ormondroyd:2026} for a Bayesian analysis favoring such a systematic in Dovekie).

To analyse the parallel direction, we convert the \Dpar\ measurements into estimates of $\omega_m$ by calibrating 
against the P-ACT chains at each redshift as discussed in \cite{Efstathiou:2025}: 
\begin{equation}
\omega_m = 0.1425 + \beta(z)({\cal D}^{\rm par} - {\cal D}^{\rm par}_{\rm P-ACT}), \label{equ:omega0}
\end{equation}
where the numerical term is the central value of $\omega_m$ from the P-ACT chains.
Values of $\beta(z)$ are given in Table~\ref{tab:parallel} in the Appendix. Since the data points
for DESI DR1 and DR2 are independent, it is straightforward to compute the best
fit value of $\omega_m$ for these data. For the BAO+SN data combinations, the data points are highly correlated. For these, we compute a covariance matrix for \Dperp\ and \Dpar\ from the Multinest chains
which is singular, reflecting the strong constraint of Eq.~\ref{equ:Hz3}. However,
the covariance structure is accurately described by four eigenvectors, which we
use to construct an inverse covariance matrix. The best fitting values of $\omega_m$ are as follows:
\begin{subequations}
\begin{eqnarray}
{\rm  DR1:} \quad \omega_m & =&  0.13857 \pm 0.00137, \label{equ:omega1} \qquad\\
{\rm DR2:} \quad \omega_m & =&  0.13950 \pm 0.00075, \label{equ:omega2} \qquad \\
{\rm  DR2}^\alpha+{\rm Unite:} \quad \omega_m & =&  0.13999 \pm 0.00055. \label{equ:omega3}  \qquad 
\end{eqnarray}
\end{subequations}
The value in Eq.~\ref{equ:omega3} is insensitive to whether we use Unite or Dovekie.  These values  differ from the 
P-ACT value $\omega_m = 0.1425 \pm 0.0012$  by  $\sim 2\sigma$. The highly 
constraining combination of DR2$^\alpha$+Unite gives a measure of $\omega_m$ 
which is in close agreement, to within plausible  sampling fluctuations,  with the predictions of the P-ACT \LCDM. However, we argue below that the interpretation is more complex.

\begin{figure}[t!]
    \centering
{\hspace{-0.4truein}\includegraphics[width=0.48\textwidth]{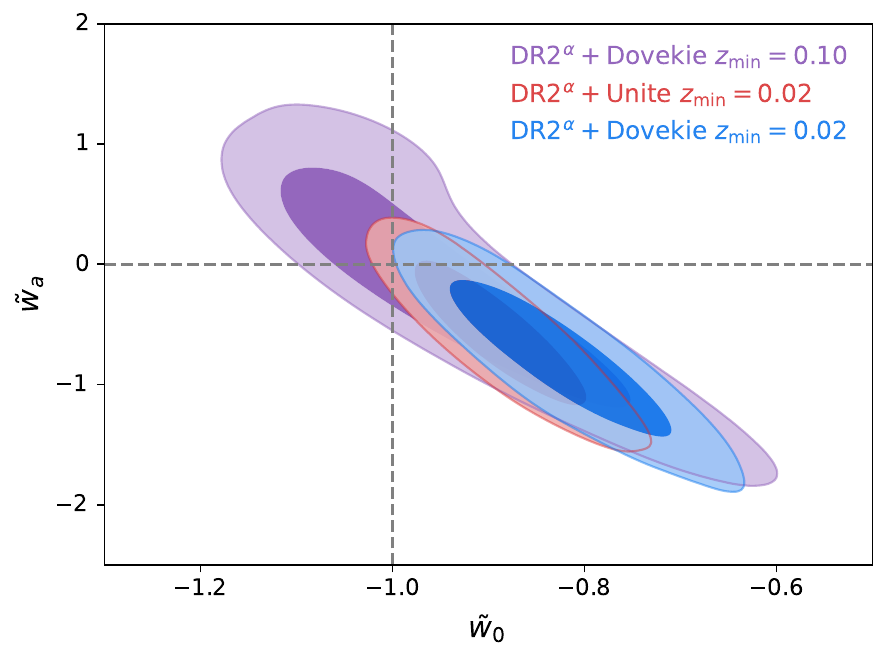}}
    \caption{$1\sigma$ and $2\sigma$ constraints on the parameters $\tilde w_0$ and $\tilde w_a$ inferred from DR2$^\alpha$+SN data.}
    \label{fig:SN_w0wa}
\end{figure}

 Figure~\ref{fig:SN_w0wa} shows constraints on
$\tilde w_0$ and $\tilde w_a$ for the three SN cases analyzed in this paper. Computing the difference in $\chi^2$ from the
best fit values fitting a \LCDM\ cosmology and the CPL parameterization, we find $\Delta \chi^2 = 4.4$ (DR2$^\alpha$+Unite),  $\Delta \chi^2 = 6.1$ (DR2$^\alpha$+Dovekie $z_{\rm min} =  0.02$) and $\Delta \chi^2 = 0.4$ (DR2$^\alpha$+Dovekie $z_{\rm min} =  0.1$)  translating to approximately $1.6\sigma$,  $2.0\sigma$ and $0.25  \sigma$ preferences for evolving dark energy given the two additional degrees of freedom of the CPL parameterization. Of course, to infer that  these numbers provide evidence for dynamical dark energy, one has to have confidence that the SN samples
are free of systematic errors over the entire redshift range  $0.02- 1$.   The additional functional freedom of the CPL parameterization is likely responding to features in the data responsible for  the deviations in \Dperp\ at $z=0.295$ noted in Fig.~\ref{fig:Main_fig}. There is certainly no good reason to favor the Dovekie compilation over the larger Unite sample, given its reliance on the Foundation SN to anchor the magnitude redshift relation at $z<0.1$.

In assessing the evidence for dynamical dark energy, it is  necessary to consider systematic errors in the analysis of the CMB. There are at least three potential sources of error: (i) systematic errors in the reionization optical depth, $\tau$, which is the least well determined parameter in the \LCDM\ cosmology; (ii) deviations from a pure power law fluctuation spectrum at large scales; (iii) systematic errors in CMB likelihoods. We briefly discuss each of these points.

\begin{enumerate}[label=(\roman*)]
\item  Motivated by the upper limits on a neutrino mass inferred from the CMB and BAO, which suggests a $\sim 3\sigma$ tension with the normal hierarchy limit of $m_\nu = 0.06\ {\rm eV}$, several authors \cite{Sailer:2026, Craig:2024, Elbers:2025,  Jhaveri:2025, Sullivan:2026} have noted the sensitivity of this result to the reionization optical depth $\tau$. The neutrino tension can be largely alleviated  if $\tau \sim 0.09$. This is much larger than the \Planck\ low multipole polarization constraint  $\tau = 0.0566^{+0.0053}_{-0.0062}$ \cite{Pagano:2020} used in the P-ACT likelihood. A high value of $\tau$ also leads to a reduction in the CMB value
of $\omega_m$. For example, \cite{Sullivan:2026} 
use a P-ACT 
likelihood similar to that used here, but replace  low multipole polarization data with a Gaussian prior $\tau=0.11 \pm 0.006$. The best fit value of $\omega_m$ changes from $\omega_m = 0.1424 \pm 0.001$ to $\omega_m = 0.1396 \pm 0.0010$,
consistent with the DR2$^\alpha+$Unite value of Eq.~\ref{equ:omega3}. Such a high value of $\tau$ is difficult to reconcile with the end-to-end simulations
used to calibrate the noise properties and systematics of the \Planck\ polarization  maps \cite{Delouis:2019, deBelsunce:2021}.  Nevertheless, the \Planck\ measurement of $\tau$ (and especially
the error) are `soft' numbers which may be affected by systematic errors which are difficult to quantify. In addition,  $\tau$ can be increased by injecting  ultra-violet photons at high redshift
leading to a low level of ionization \cite{Weiner:2026}. Such models are difficult to test with current CMB data.
    
\item The first results from WMAP revealed evidence for `anomalies' in the TT power spectrum at low multipoles, which may indicate deviations from
a pure power law curvature fluctuation spectrum \cite{Peiris:2003}. These
features were reproduced with high accuracy by \Planck\ \cite{Planck_2013}. The statistical significance of these deviations  (and other low multipole anomalies)
is difficult to quantify, but lies somewhere in the range $\sim 2-3\sigma$. In cosmological parameter estimation, these features  are treated as statistical fluctuations, but if power  is suppressed as  a consequence of physical processes, the inferred value of $\tau$ will be biased since at low multipoles the EE power spectrum scales as $C_\ell^{\rm EE} \sim \tau^2 {\cal P}(k \sim \ell/D_M (z_{\rm rei}))$, where ${\cal P}$ is the curvature power spectrum and $z_{\rm rei}$ is the redshift of reionization  \cite{Jhaveri:2026}. The interpretation of cosmological  parameters from the CMB is therefore conditional on the
assumption that the curvature power spectrum is featureless,
described by an amplitude and a scalar spectral index.

\item Until recently, CMB likelihoods were statistically
dominated by Planck. The Planck spectra and
likelihoods have been tested extensively and the
parameters of the ΛCDM cosmology are usually
found to be consistent to  within a fraction of a
standard deviation \cite{Plancklikelihood_2015, Planck:2018, Efstathiou:2021, Jense:2026}, provided the reionization optical depth is constrained to a value
consistent with Planck SROLL2. The 
analysis reported in \cite{SPT:2023}, on the other hand, uses
several different likelihoods some of which lead to different conclusions. We have repeated the analysis described above using \LCDM\ chains based on two
likelihoods constructed by the SPT collaboration \cite{SPT:2023} which we label SPA and SPAnotau\footnote{The chains can be found at~https://lambda.gsfc.nasa.gov/product/spt.\newline S1920lite\_MPP\_PACT\_PR4lens\_actdr6lens\_No\_OLE~(SPA);~S1920lite\_MPP\_PACT\_\newline PR4lens\_actdr6lens\_ \newline no\_tau\_prior\_No\_OLE (SPAnotau). SPA is a combination of \Planck, ACT and SPT power spectra with \Planck,  ACT DR6 and SPT3G~\cite{Ge:2025} lensing and an SROLL2 like Gaussian prior
on the reionization optical depth  $\tau = 0.051 \pm 0.006$. The SPAnotau chains replace the Gaussian prior on $\tau$ with a uniform prior between $\tau = 0.01-0.4$}. To  high accuracy, the tension between these likelihoods and Eqs~\ref{equ:omega1}-\ref{equ:omega3} can be quantified by comparing the marginalized values of $\omega_m$: $\omega_m = 0.14332 \pm 0.00091$ for SPA resulting in a tension
with DR2$\alpha$+Unite of $3.1\sigma$; $\omega_m = 0.1417 \pm 0.00124$ for SPAnotau reducing the tension to $1.3\sigma$. It is the high value of $\omega_m$ for the SPA likelihood that leads to
the conclusions of \cite{SPT:2026} of an increased tension between CMB measurements and DESI BAO. This tension is relaxed using SPAnotau, which gives a value of $\tau = 0.076 \pm 0.013$ constrained largely from CMB lensing.

\end{enumerate}

As cosmological measurements have become more precise, the subject has entered  a phase in which small systematics can be mistaken
for important new physics such as evolving dark energy. One needs to be certain of the photometric matching of SN to an accuracy of $\sim 0.01\ {\rm mag.}$ over the entire redshift range $z = 0.02-1$  (see Fig.~\ref{fig:Magres}) 
before concluding definitively that the \LCDM\ cosmology has failed. In addition, there are a number of potential systematic effects in the CMB analyses
that can affect cosmological inferences. It is not clear whether the  \LCDM\  parameters derived from  different combinations of CMB data (point (iii))  are caused by statistical fluctuations or small systematic errors in the data at high multipoles (including CMB lensing).  Points (i) and (ii) both involve potential systematics at large angular scales. The simplest way to mitigate against these is to discard CMB data at multipoles $\simlt 50$, relying on CMB lensing to constrain $\tau$ and accepting the loss of precision in cosmological parameters.

The goal for the future is to progress from precise tests of the \LCDM\ cosmology to accurate ones, free of systematics. Until small systematics are better understood,  there is no
compelling case either  to reject the \LCDM\ cosmology or to conclude that
dark energy is evolving.

\begin{acknowledgments}
I thank Simon White, Tamara Davis and colleagues for comments on the manuscript. I am very grateful to  the Unite team for sharing their data prior to publication. 
This paper has intentionally used chains in the public domain so that the results can easily be reproduced. Claude was used
to make figures.

\end{acknowledgments}
\vspace{-0.2 truein}

\bibliography{biblio.bib}

\clearpage
\onecolumngrid
\appendix*

\section{END MATTER}
\label{sec:appA}

  This appendix provides tables of relevant parameters used in the figures
  and text of the main paper. We give fitted parameters only for the   DR2$^\alpha$+Unite analysis. As discussed in the main text, the parameters for DR2$^\alpha$+Dovekie ($z_{\rm min} = 0.02$) and BAO+Dovekie ($z_{\rm min} = 0.1$) are almost identical to those of DR2$^\alpha$+Unite, except for the BGS bin (see Fig.~\ref{fig:Main_fig}). 

\begin{table*}[h]

\begin{center}

\caption{The first column lists the DESI sample and the second column lists the effective redshift. The third column gives the rotation angles $\gamma$ in radians used
  to define \Dpar\ and \Dperp. The fourth column gives the values of \Dperp\ computed from the P-ACT \LCDM\ chains. These numbers have negligible error.  Columns 5 and 6 list measurements of \Dperp\ from DESI DR2 and DR2$^\alpha$+Unite  together with $\pm 1\sigma$ errors. The columns labelled $N_\sigma$ give the difference between these values  and  the P-ACT values  in units of DESI standard deviation. The last two
  columns list $D_M/r_d$ and $D_H/r_d$ for DR2$^\alpha$+Unite for comparison with future DESI data  releases.  }

\label{tab:perp}
\resizebox{\textwidth}{!}{
\begin{tabular}{l|c|c|c|c|c|c|c|c| c} \hline 
    &  &  &  P-ACT    & \multispan2{\hspace{-0.0 truein} DR2} & \multispan4{\hspace{0.1 truein} {\hspace{1.05 truein} ${\rm DR2}^\alpha+{\rm Unite}$}}  \hspace{1.3 truein}\\ 
    &  &  &          & \multispan2\quad$\overbrace{\hspace{0.95truein}}$ & \multispan4$\overbrace{\hspace{3.6truein}}$  \\
    sample  &  $z_{\rm eff}$ & $\gamma$ & ${\cal D}^{\rm perp}$ & 
    ${\cal D}^{\rm perp}$ & $N_\sigma$ & ${\cal D}^{\rm perp}$ & $N_\sigma$ &  $D_M/r_d$ & $D_H/r_d$  \\\hline
    BGS & $0.295$ & \ $1.190$ \ & \ $1.936$ \ & -  & -  &\ $1.889 \pm 0.021$\ & \ \ $2.2$\  &\ \ $8.122 \pm 0.043$ \  &\  $25.339 \pm 0.123$ \\  
 LRG1 & \  $0.510$ \ &  \ $0.816$ \ & \  $5.738$ \ & $5.07\pm 0.36$ & \  \ $1.9$ \ & \ \ $5.704 \pm 0.046$ \ & \ \ $0.74$ \ &\  $13.254 \pm 0.059$  \ & $22.438 \pm 0.075$  \\ 
 LRG2  & $0.706$ &\  $0.499$ \ & \    $9.247$ \ & $8.78\pm0.34$ &\ \  $1.4$ \ &  \ \ $9.247\pm 0.077$ \ & \ \  $0.0$ \ & \ $17.411 \pm 0.063$ \  & $20.025 \pm 0.085$ \\
 LRG3+ELG1 & $0.934$ &  $0.241$\ &\  $11.832$ \ & $11.98 \pm 0.21$ &\ $-0.70$ \ & $11.853 \pm 0.089$ \ & \ $-0.24$ \ \  & \ $21.686\pm 0.066$ \  & $17.529 \pm 0.093$ \\
 ELG2 & $1.321$ & \ \   $0.023$ \  & $13.440$\  & $13.54  \pm 0.22$ & \ $-0.45$ \   & $13.440\pm 0.075$ \ &\  $0.0$\ \   & \ $27.771 \pm 0.075$ \ & $14.078 \pm 0.076$  \\  
 QSO & $1.484$ & \  $-0.023$ \ \  &\  $13.582$ \ &  $13.50 \pm 0.50$ & \  \ $0.26$ \  &  $13.569 \pm 0.070$ \ & \ $0.18$\ \  & \ $29.967 \pm 0.080$ \ & $12.894 \pm 0.079$ \\
 Ly$\alpha$ & $2.33$ \ & \  $-0.106$\ \ & \ $12.704$ \ &  $12.627 \pm 0.08$  & \ \  $0.96$ \ &  $12.634 \pm 0.068$ \ &\ \  $1.0$ \  & \  $38.875  \pm 0.106$ \ &  $8.603\pm 0.062$    \\ \hline
\end{tabular}}
\end{center}
\end{table*}

\begin{table*}[h]¯

\begin{center}

  \caption{The first two columns are as in Table~\ref{tab:perp}. The third column gives the slope $\beta$ of Eq.~\ref{equ:omega0} at each redshift necessary to convert measurements of \Dpar\ into constraints on $\omega_m$ according to the \LCDM\ cosmology. The columns labeled   list the values of  \Dpar\ conditional on \Planck\ \LCDM, for DESI DR2 and DR2$^\alpha$ together with $\pm 1\sigma$ errors.  The columns labeled $\omega_m$ convert the measurements of ${\cal D}^{\rm par}_{\rm cond}$ into measurements of $\omega_m$ using Eq.~\ref{equ:omega0}.}

\label{tab:parallel}

\smallskip
\resizebox{\textwidth}{!}{
\begin{tabular}{l|c|c|c|c|c|c|c} \hline 
    &  &  &  P-ACT &      \multispan2{\hspace{-0.0 truein} DR2} & \multispan2{\hspace{0.28 truein} { ${\rm DR2}^\alpha+{\rm Unite}$} \hspace{0.23truein}}  \\ 
    &  &  &            & \multispan2$\quad\overbrace{\hspace{1.8 truein}}$ & \multispan2$\quad\overbrace{\hspace{2.1truein}}$  \\
    sample  &  $z_{\rm eff}$ & $100\beta$ & ${\cal D}^{\rm par}$ & ${\cal D}^{\rm par}_{\rm cond}$&   $\omega_m$  &   ${\cal D}^{\rm par}_{\rm cond}$ &  $\omega_m$  \\ \hline
BGS        &\ $0.295$ \ &\ $0.675$ \ & \ $27.03 \pm 0.17$ \ &     $-$              &       $-$                & \ $26.666 \pm 0.116$ \ &$0.14002 \pm 0.00078$ \\
LRG1       &\ $0.510$ \ &\ $0.870$ \ & \ $25.78 \pm 0.13$ \ & \ $25.62 \pm 0.17$ \ & \  $0.1411 \pm 0.0015$ \ & \ $25.443 \pm 0.081$ \ & $0.13954 \pm 0.00070$   \\
LRG2       &\ $0.706$ \ &\ $0.958$ \ & \ $25.17 \pm 0.12$ \ & \ $24.61 \pm 0.16$ \ & \  $0.1371 \pm 0.0015$ \ & \ $24.872 \pm 0.069$ \ & $0.13963 \pm 0.00066$  \\
LRG3+ELG1  &\ $0.934$ \ &\ $0.972$ \ & \ $25.52 \pm 0.12$ \ & \ $25.18 \pm 0.13$ \ & \  $0.1388 \pm 0.0013$ \ & \ $25.234 \pm 0.068$ \ & $0.13939 \pm 0.00066$ \\
ELG2       &\ $1.321$ \ &\ $0.955$ \ & \ $28.37 \pm 0.12$ \ & \ $27.98 \pm 0.28$ \ & \  $0.1387\pm 0.0027$ \ & \ $28.085 \pm 0.069$ \ & $0.13976 \pm 0.00066$   \\
QSO        &\ $1.484$ \ &\ $0.954$ \ & \ $29.95 \pm 0.12$ \ & \ $30.15 \pm 0.67$ \ & \  $0.1444 \pm 0.0064$ \ & \ $29.673 \pm 0.070$ \ & $0.13984 \pm 0.00067$  \\
Ly$\alpha$ &\ $2.33$\ \ &\ $1.007$ \ & \ $38.05 \pm 0.11$ \ & \ $38.25 \pm 0.26$ \ & \  $0.1445 \pm 0.0027$ \ & \ $37.816 \pm 0.083$ \ &  $0.14012 \pm 0.00083$  \\\hline
\end{tabular}}
\end{center}

\end{table*}

\end{document}